\documentclass[preprint]{aastex}

\usepackage{epsf}
\usepackage{emulateapj5}
\usepackage{onecolfloat}
\usepackage{apjfonts}
\usepackage{amsmath}

\newcommand{\simgt}{\,\rlap{\lower 3.5 pt \hbox{$\mathchar \sim$}} \raise
1pt \hbox {$>$}\,}
\newcommand{\simlt}{\,\rlap{\lower 3.5 pt \hbox{$\mathchar \sim$}} \raise
1pt \hbox {$<$}\,}

\begin{document}
\submitted{The Astrophysical Journal, accepted}
\vspace{1mm}
\slugcomment{{\em The Astrophysical Journal Letters in press}}

\shorttitle{DM Structures Not That Trivial}
\shortauthors{Schmidt, Hansen \& Macci\`o}

\twocolumn[
\lefthead{dark matter structures were not that trivial}
\righthead{Schmidt, Hansen \& Macci\`o}

\title{Alas, the dark matter structures were not that trivial}

\author{Kasper B. Schmidt\altaffilmark{1,2}, Steen H. Hansen\altaffilmark{1}, \& Andrea V. Macci\`o\altaffilmark{2}}

\begin{abstract}
The radial density profile of dark matter structures has been observed
to have an almost universal behaviour in numerical simulations,
however, the physical reason for this behaviour remains unclear.
It has previously been shown that if the
pseudo phase-space density,
$\rho/\sigma_d^\epsilon$, is a beautifully simple power-law in radius,
with the {\em ``golden values''} $\epsilon=3$ and $d=r$ (i.e., the phase-space density is only dependent on the radial component of the velocity dispersion), then
one can analytically derive the radial variation of the
mass profile, dispersion profile etc.
That would imply, if correct, that we just have to explain
why $\rho/\sigma^3_r \sim r^{-\alpha}$, and then we would
understand everything about equilibrated DM structures.
Here we use a set of simulated galaxies and clusters
of galaxies to demonstrate that there are no such golden
values, but that each structure instead has its own set of values. Considering the same structure at different redshifts shows no evolution of the phase-space parameters towards fixed points. There is also no clear connection between the halo virialized mass and these parameters.
This implies that we still do not understand the origin of
the profiles of dark matter structures.
\end{abstract}


]

\altaffiltext{1}{Dark Cosmology Centre, Niels Bohr Institute, University of Copenhagen,\\
Juliane Maries Vej 30, 2100 Copenhagen, Denmark}
\altaffiltext{2}{Max Planck Institut f\"ur Astronomie, K\"onigstuhl 17, 69117 Heidelberg, Germany}


\section{Introduction}
\label{sec:intro}

According to numerical
simulations of dark matter (DM) structures,
the mass density profile, $\rho(r)$,
changes from something with a fairly shallow profile in the central
region, $\gamma \equiv d{\rm ln}\rho/d{\rm ln}r \sim -1$ (or maybe
zero), to something steeper in the outer region, $\gamma \sim -3$ (or
maybe steeper) \citep{nfw,moore,diemand} (see also
\cite{reed,stoehr,navarro2004,graham,merritt,ascasibar}).  For the
largest structures, like galaxy clusters, there appears to be fair
agreement between numerical predictions and observations
concerning the central steepness
\citep{pointe,sand,buote,broadhurst,vikhlinin}, however, for smaller
structures, like galaxies or dwarf galaxies, observations tend to
indicate central cores~\citep{salucci,gilmore,wilkinson}.
Few purely theoretical attempts have been made to understand
the origin of this density profile, e.g.
\cite{gsmh,henriksen}, with varying level of success.

A completely different approach is to search for simple phenomenological
relations in the numerical simulations, such as finding straight
lines in some parameter space. The idea is then that such
phenomenological relations may reduce the complexity of the
Jeans equation, which can then be solved analytically.

One of the most successful attempts in this direction was
sparked by the discovery that the pseudo phase-space density
is approximately a power-law in radius, $\rho/\sigma_r^3 \sim r^{-\alpha}$ 
(Taylor \& Navarro 2001).
The most simple analytical solutions to
this problem showed, that the density slopes could vary in
the range from -1 to -3 \citep{hansenjeans}, in excellent
agreement with numerical results of \cite{nfw}.
The analytical investigations were taken to a higher level
in \cite{austin}, where it was demonstrated that there is a
characteristic value $\alpha=1.944$ when one considers
isotropic structures. Shortly after \cite{dehnenmclaughlin}
used the results of numerical simulations \citep{diemand04a, diemand04b}
to show that the ``golden values'' $\alpha=1.944$ and $\epsilon=3$
indeed provides
a very good fit, when one is using the {\em radial} velocity
dispersion in the pseudo phase-space density.
Dehnen \& McLaughlin (2005, henceforth DM05) also solved the Jeans equation under
this assumption, and demonstrated explicitly that one hereby
can derive analytically all relevant profiles for the DM structure. Many other
authors have considered similar pseudo phase-space densities,
e.g. \cite{hansenzemp,knebe06,stadel08, knollmann08}, \cite{ascasibar,zait08,hese08,navarro2008,lapi08}.

All this implies, that if we can explain the origin of the
very simple connection,
$\rho/\sigma_r^3 \sim r^{-\alpha}$, then we have complete
understanding of the DM structures. However, this
is under the implicit assumption that the 3 golden values
 are indeed the same for all structures,
namely that $\alpha = 1.944$, and that the relevant quantity
to consider is the radial dispersion, $\sigma_d^\epsilon$
with $d=r$ and $\epsilon = 3$.

We will here use the results of recent numerical simulations
to demonstrate that this is {\em not} the case, and that there is
{\em no} simple universal pseudo phase-space density for equilibrated
DM structures. Given our findings it therefore appears that few
theoretical approaches that successfully explain the origin of
the cosmological profiles such as the Barcelona model \citep{man03,gsmh} remain.


\section{Generalized Pseudo Phase-Space Density}\label{sec:psd}

In order to test whether a generalized phase-space density exists, we consider the relation 
\begin{equation}\label{eq:psd}
\frac{\rho}{\sigma_d^\epsilon} \propto r^{-\alpha}	\; .
\end{equation}
Here we have defined the general velocity dispersion as \citep{bohansen,schmidt}
\begin{equation}\label{eq:sigd}
\sigma_d^\epsilon = \sigma_r^\epsilon\left(1+D\beta\right)^{\epsilon/2} \; .
\end{equation}
Here $\beta(r)=1-\frac{\sigma_{tan}^2}{\sigma_{rad}^2}$ is the usual velocity anisotropy parameter, where $\sigma_\textrm{rad}$ 
and $\sigma_\textrm{tan}$ is the radial and tangential component of the velocity dispersion respectively.
Thus setting $D=0$, corresponds to using the radial component of the DM structure velocity dispersion in the phase-space 
density expression. Allowing $D\neq 0$ the phase-space density depends on a velocity dispersion which can be any 
combination of $\sigma_\textrm{rad}$ and $\sigma_\textrm{tan}$, e.g. $D=-\frac{2}{3}$ corresponds to using 
$\sigma_\textrm{tot}$, and $D=1$ corresponds to $\sigma_\textrm{tan}$.

Here we use the rather simple analytical pseudo phase-space density, however the actual 6 dimensional phase-space density is different from this one and is not a power-law in radius according to simulations \citep{stadel08}. 

\section{Numerical Simulations}\label{sec:sim}

To test if the values of $D$, $\alpha$ and $\epsilon$ are the same for all structures, we used a set of intermediate and high resolution simulated DM structures. These structures are all created using the PKDGRAV treecode by Joachim Stadel and Thomas Quinn \citep{stadel}. 
The high resolution simulation  'Via Lactea' includes one highly equilibrated structure of mass $M_{halo}
= 1.77 \times 10^{12} M_\odot$, containing about 84 million particles \citep{diemand}. This structure did not experience any major mergers since $z=1$ and all the quantities are extracted in spherical bins. 
The rest of the structures are galaxy-size and cluster-size DM halos based on either a WMAP 1 year or a WMAP 3 year cosmology.
The initial conditions for these structures are generated with the GRAFIC2 package \citep{bert01}. The starting  redshifts $z_i$  are set to  the time when  the standard
deviation  of the  smallest density  fluctuations resolved  within the simulation box  reaches $0.2$ (the smallest scale  resolved within the initial conditions is defined as twice the intra-particle distance).
All the halos were identified using a spherical overdensity algorithm \citep{mac07}.
The cluster-like halos have been extracted from a 63.9 $\textrm{Mpc}/h$ simulation containing $600^3$ particles, with a mass resolution of $m_p=8.98 \times 10^7 M_{\odot}/h$. The masses of the clusters used for 
this study are 2.1, 1.8, and 1.6 $\times 10^{14} M_{\odot}/h$. The galaxy-size halos have been obtained by re-simulating halos found in the previous simulation at high resolution. The simulated halos 
are in the mass range $0.9-2.5 \times 10^{12} M_{\odot}/h$ and have a mass resolution of $m_p=4.16 \times 10^5 M_{\odot}/h$. That gives a minimum number of particles per halo of about $2.5 \times 10^6$. The high resolution cluster C$_{HR}$.W3 has 11 million particles within its virial radius and a mass of $ M=1.81\times 10^{14} M_{\odot}/h$.

From these numerical simulations we directly calculate all the relevant quanteties, such as $\rho(r)$, $\sigma_r(r)$, $\sigma_\theta(r)$, $\sigma_\phi(r)$ where the $\sigma$'s are combined to obtain $\beta(r)$.  These profiles can then be compared to the pseudo phase-space density defined in Eq.~(\ref{eq:psd}).


\section{Monte Carlo Code}\label{sec:MC}

In order to test whether the suggested golden values of DM05 and \cite{austin} do indeed exist, we wrote a Monte Carlo (MC) code to optimize the parameters of the phase-space density in Eq.~(\ref{eq:psd}), for each of the simulated DM structures. 

The MC code is based on the temperature annealing principle \citep{kirk, hansenmc}. 
We want to optimize the parameter set $(D,\alpha,\epsilon)$, so that it makes the LHS and RHS of relation~(\ref{eq:psd}) converge towards the expected power-law relation. In order to do that we search the parameter space and for each jump estimate the $\chi^2$ value of the relation defined by
\begin{equation}\label{eq:c2}
\chi^2 = \Sigma_i \left(\frac{f_1(x_i) - f_2(x_i)}{df_2(x_i)}\right)^2	\;.
\end{equation} 
Here $x_i$ corresponds to the data input from the simulations, $f_1$ and $f_2$ corresponds to the LHS and RHS of the relation and $df_2$ is the error on $f_2$. Since we are dealing with simulations we have no reasonable estimate of the error $df_2$. Therefore we use $df_2=0.05\frac{\rho}{\sigma_r^3}$. Choosing different kinds of errors (e.g. $\frac{\rho}{\sigma_D^\epsilon}$, $\frac{\rho}{\sigma_D^3}$, $\frac{\rho}{\sigma_r^\epsilon}$ and $r^{-\alpha}$) with different magnitudes (0.05, 0.07 and 0.10) has no significant systematic effect on the final result.
Since there is the possibility of local minima in the '$\chi^2$-landscape' we have implemented the metropolis choice in our code \citep{metropolis}. All technical details of this code can be found in \cite{schmidt}.

\begin{table}
\caption{Phase-Space Density fitting parameters. The case  marked with an *
is the one used in \cite{hansenjeans,austin,dehnenmclaughlin}.}
\begin{tabular}[c]{|c|c|c|c|}
\hline
$D$ & Phase-Space Density & $\alpha$ & $\epsilon$  \\
\hline
\hline
1 &  ${\rho}/\left( {\frac{1}{2}\left(\sigma_\phi^2+\sigma_\theta^2\right)^{\epsilon/2}} \right)  $ & $2.13\pm0.03$  & $4.12\pm0.47$\\
0*&  ${\rho}/{\sigma_r^\epsilon}$ & $1.94\pm0.02$  & $3.15\pm0.29$\\
-1&  ${\rho}/ {\left(2\sigma_r^2-\frac{1}{2}\left(\sigma_\phi^2+\sigma_\theta^2\right)\right)^{\epsilon/2} } $ & $1.75\pm0.03$  &$2.18\pm0.47$ \\
$-\frac{2}{3}$ 	&  ${\rho}/{\left(\frac{1}{3}\left(\sigma_r^2+\sigma_\phi^2+\sigma_\theta^2\right)\right)^{\epsilon/2} }$ & $1.81\pm0.02$ & $2.50\pm0.38$ \\
\hline
\end{tabular}
\label{tab:alpeps}
\end{table}


\section{Results}\label{sec:results}

\begin{figure}[t]
\centerline{\epsfxsize=3.2in \epsffile{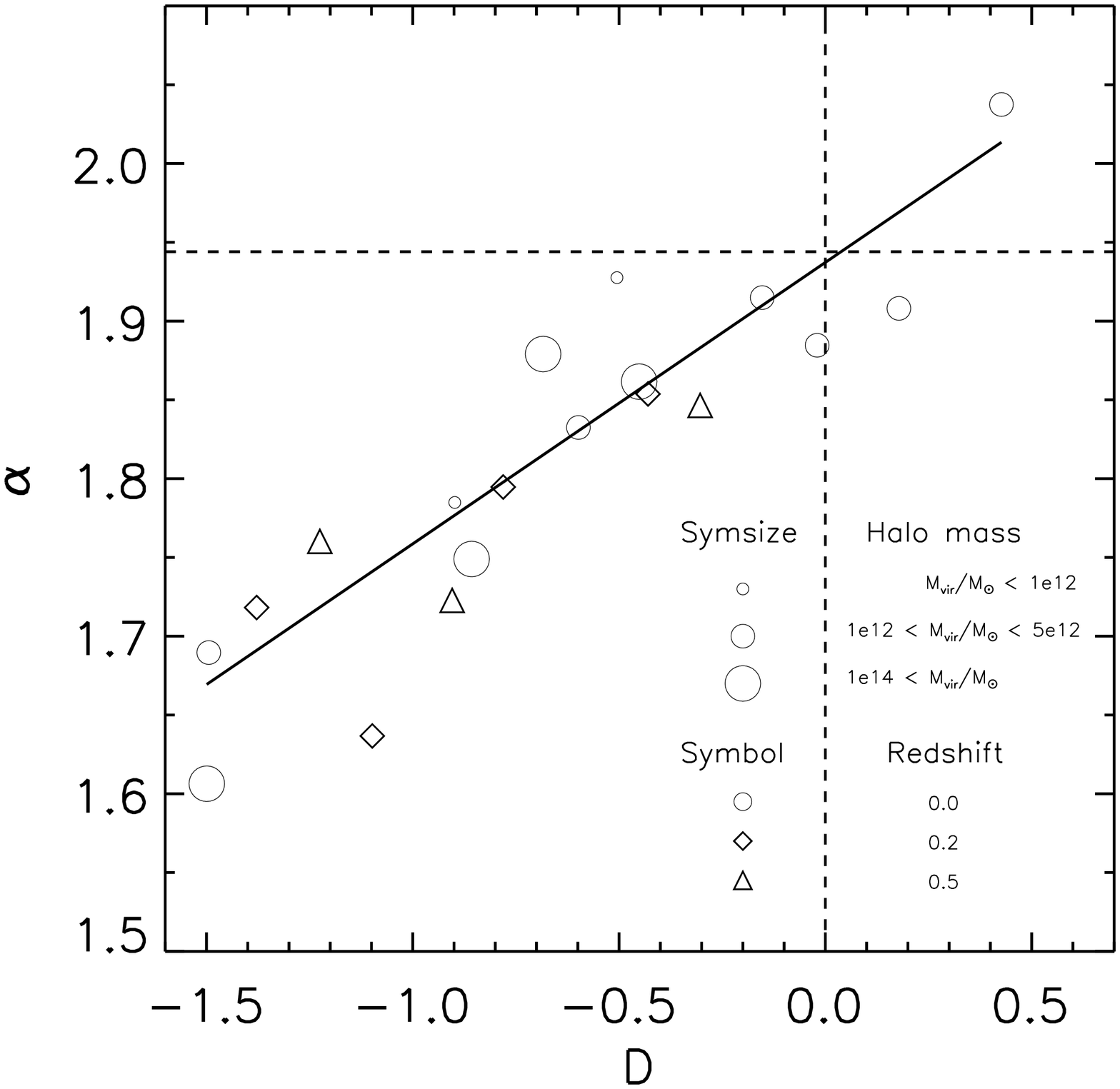}}
\caption{The optimized $D$ and $\alpha$ values from the MC code plotted against each other. The vertical and horizontal dashed lines indicate the golden values suggested by \cite{austin} and \cite{dehnenmclaughlin}. The solid line is a linear fit to the results for the $z=0$ structures (circles) and is given by $\alpha = (0.19 \pm 0.02)\times D + (1.94\pm 0.02)$. This indicates that the suggested golden values are just a consequence of using $\sigma_r$ in the phase-space density relation. The results for the galaxy-size WMAP1 structures at $z=0.2$ (diamonds) and $z=0.5$ (triangles) are over-plotted for comparison, and shows a trend very similar to the $z=0$ case. No correlation between halo mass (indicated by different symbol size) and the fitting parameters is detected.}
\label{fig:rpbest}
\end{figure}

Combining the simulated DM structures with the MC code, we are able to estimate the parameters that optimizes the phase-space density relation for each structure. If there should be a general phase-space density relation, each structure should have the same optimized parameters. We see in Fig.~\ref{fig:rpbest} that this is not the case. In Fig.~\ref{fig:rpbest} we have indicated the suggested golden values as horizontal and vertical dashed lines.

\begin{figure}[t]
\centerline{\epsfxsize=3.2in \epsffile{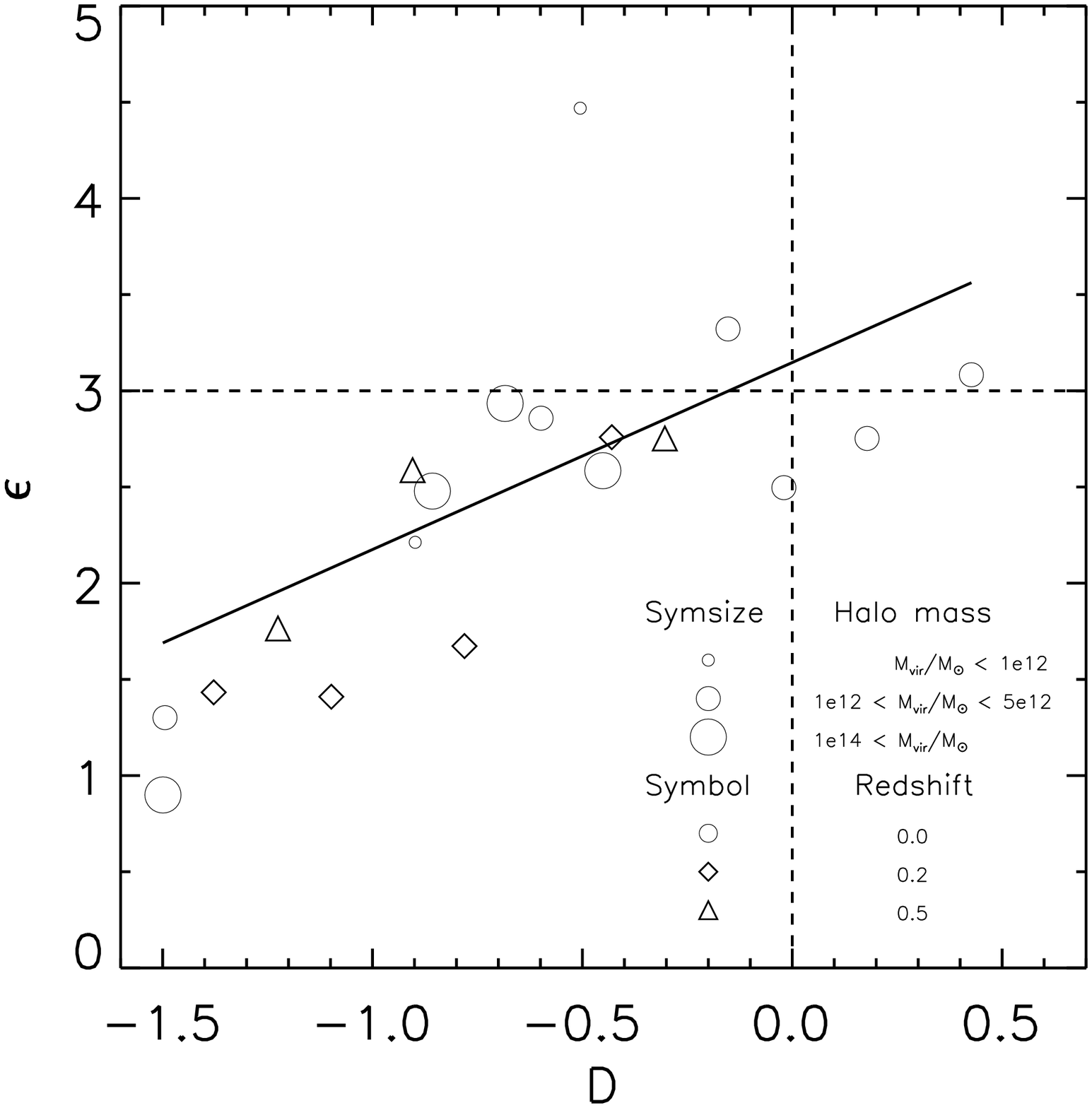}}
\caption{The same figure as Fig.~\ref{fig:rpbest} for $D$ and $\epsilon$ with a linear fit given by $\epsilon = (0.97 \pm 0.37)\times D + (3.15\pm0.29)$.}
\label{fig:rpbest2}
\end{figure}

The obtained (roughly) linear relations in Figs.~\ref{fig:rpbest} and \ref{fig:rpbest2} are
\begin{eqnarray}
\alpha &=& (0.19 \pm 0.02)\times D + (1.94\pm 0.02) \label{eq:alpdrel}\\
\epsilon &=& (0.97 \pm 0.37)\times D + (3.15\pm0.29) \label{eq:epsdrel} 	\; .
\end{eqnarray}
This shows that a generalized phase-space density relation does not exist.
Thus our results suggest that the hunt for a physical explanation of the (often assumed universal) power-law appearance of $\rho/\sigma^3$, is probably a dead end. It seems that this expression is nothing more than a possible fitting function with the nice property of having the same physical units as the phase space density.

If we force $D=0$ (like DM05) we get from the relations~(\ref{eq:alpdrel}) and (\ref{eq:epsdrel}) that $\alpha=1.94\pm 0.02$ and $\epsilon=3.15\pm0.29$. These values are in excellent agreement with the results from DM05. If we use other values of $D$, i.e., phase-space densities with combinations of the different velocity dispersion components we get the parameter values listed in table \ref{tab:alpeps}.

\begin{figure}[t]
\centerline{\epsfxsize=3.2in \epsffile{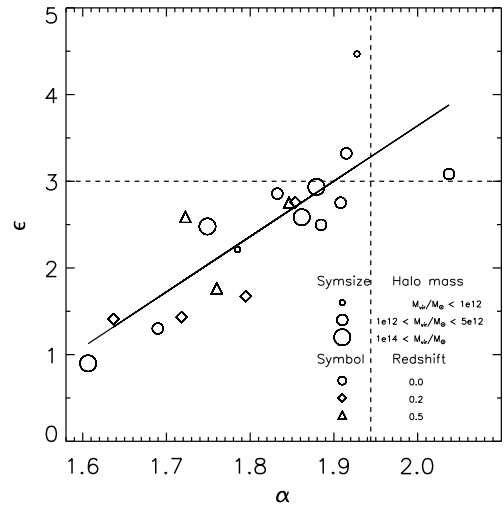}}
\caption{The same figure as Fig.~\ref{fig:rpbest} for $\alpha$ and $\epsilon$ with a linear fit given by $\epsilon = (6.39 \pm 1.44)\times \alpha + (-9.14 \pm2.65)$.   }
\label{fig:rpbest3}
\end{figure}

Knowing that the parameters that optimizes the phase-space density relation in Eq.~(\ref{eq:psd}) for simulated structures at redshift $z=0$ are related, it would be interesting to see whether such relations are also present at higher redshifts or not. We use $z > 0$ snapshots for the WMAP1 galaxy-size structures (clusters are not significantly relaxed at high redshifts). Running the MC code with these DM halos gave results very similar to the ones for the $z=0$ structures. In general we did not find any indication of a redshift dependence for the optimized values, showing that there is no special attractor for the values of $D$, $\alpha$ and $\epsilon$. The results at $z > 0$ (triangles and diamonds) are plotted together with the $z=0$ ones in Figs.~\ref{fig:rpbest}-\ref{fig:rpbest3}. The obtained linear relations between the optimized parameters are not affected significantly by redshift, and therefore our calculations suggest that Eqs.~(\ref{eq:alpdrel}) and (\ref{eq:epsdrel}) are valid for all redshifts. 

Furthermore we find no significant correlation between the virial mass of the simulated structures and the $\alpha$, $\epsilon$ and $D$ values (see Figs.~\ref{fig:rpbest}-\ref{fig:rpbest3}).

\section{Conclusions}
Using a set of numerically simulated galaxy and cluster sized DM structures and analysing them with a Monte Carlo code, we show that no generalized pseudo phase-space density relation seems to exist in general. We have thus shown, that the previously suggested relation $\rho/\sigma^3\sim r^{-\alpha}$ does not hold universally. The redshift and mass independence of our results show that there is no special attractor for the parameters describing the generalized phase-space density.

Instead we happen to identify a set of seemingly linear relations between the parameters $D$, $\alpha$ and $\epsilon$ (describing the generalized pseudo phase-space density from Eq.~(\ref{eq:psd})), which we have parametrized in Eqs.~(\ref{eq:alpdrel}) and (\ref{eq:epsdrel}). 

Thus, given our findings that $\rho/\sigma^3$ is nothing but a
nice fitting formula and not a physical attractor, we are still far from truly understanding the density profile of DM structures.

\section*{Acknowledgments} 

It is a pleasure to thank Juerg Diemand for kindly providing part of the
numerically simulated data used in the figures. We are grateful to Jin H. An and Bo V. Hansen for fruitful discussions.
Part of the numerical simulations were performed on  the  PIA cluster  of  the
Max-Planck-Institut f\"ur Astronomie at the Rechenzentrum in Garching.
The Dark Cosmology Centre is funded by the Danish National Research Foundation.


\end{document}